# "[AI peers] are people learning from the same standpoint": Perception of AI characters in a Collaborative Science Investigation

Soo Hyoung Joo[1†][0000-0003-1899-4333] and Eunhye Grace Ko[2†][0000-0002-8900-2197]

[1] Teacher's College, Columbia University, New York, NY 10027, USA
[2] School of Information, The University of Texas at Austin TX78701, USA

† Co–first authors (these authors contributed equally to this work).
shj2118@tc.columbia.edu, eko@utexas.edu

**Abstract.** While the complexity of 21st-century demands has promoted pedagogical approaches to foster complex competencies, a persistent gap remains between in-class learning activities and individualized learning or assessment practices. To address this gap, studies have explored the use of AI-generated characters in learning and assessment. One such attempt is scenario-based assessment (SBA), a technique that not only measures but also fosters the development of competencies throughout the assessment process. SBA introduces simulated agents to provide an authentic social-interactional context allowing for the assessment of competency-based constructs while mitigating the unpredictability of real-life interactions. Recent advancements in multimodal AI, such as text-to-video technology, allow these agents to be enhanced to AI-generated characters. This explanatory mixed-method study investigates how learners perceived the AI characters taking the role of mentor and teammates in an SBA mirroring the context of performing a collaborative science investigation. Specifically, we examined the Likert scale responses of 56 high school learners' perception of the AI characters, in terms of trust, social presence, and effectiveness. We also analyzed the relationships between these factors and their impact on the intention to adopt AI characters for future learning through PLS-SEM. Our findings indicated that learners' trust shaped their sense of social presence with the AI characters, enhancing perceived effectiveness. Additionally, the qualitative analysis further highlighted factors that foster trust, such as the credibility of materials and alignment with learning goals, as well as the pivotal role of social presence in creating a collaborative context. As an early study examining the role of AI characters in learning, this research contributes to the understanding of key factors influencing engagement with multimodal AI characters adding to the growing body of research on AI-assisted education.

## 1 Introduction

The complexity of 21st-century demands has led to the development of pedagogical approaches to foster complex competencies such as collaborative problem-solving and critical thinking [1], yet a persistent gap remains between learning activities in class settings and individualized learning or assessment practices [2][3]. With the advent of Generative AI (GenAI) technology, studies have explored the use of AI-generated characters to promote the development and assessment of these competencies (e.g., [4]).

Given the novelty of this approach, investigation is needed to determine to what extent the interaction with these characters is conducive to learning, and if so, what aspects impact the quality of these interactions. Previous studies have examined various aspects of these characters, such as their voice (e.g., [5]), style of communication (e.g., [6], [7]), familiarity [5], and the credibility of the source materials, and how these factors influence learners' perceptions of AI characters [8] and, in turn, their intent to engage in further AI-driven education. However, many of these studies focus on chat-based interfaces, which—while promising—may limit inclusivity, engagement, and long-term learning effects, particularly for learners with short attention spans shaped by rapid-content platforms [9]. A shift toward multimodal AI can offer a more authentic and immersive learning experience.

One approach where multimodal AI-generated characters are increasingly adopted is scenario-based assessment (SBA), an assessment technique that attempts to both measure and foster the development of real-world competencies throughout the assessment process [10][11]. Simulated characters play a pivotal role in SBAs, driving the scenario narrative and creating a social-interactional context that necessitates the purposeful deployment of real-life competencies [12]. Traditionally, these characters were represented through static images with voice-overs, which, while effective in structuring interactions, often lacked the realism necessary to fully engage learners [13]. Recent advancements in multimodal GenAI technology offer new possibilities for enhancing SBAs by integrating AI-generated characters. For second language (L2) learners, interaction with these AI characters is critical, as they can serve as valuable scaffolds in addressing the dual challenge of mastering complex scientific concepts while communicating in a non-native language.

As part of a larger effort to develop and validate a scenario-based language assessment for science purposes (SBLA-SP) incorporating multimodal AI-generated characters, this study investigates learners' perceptions of these AI characters in terms of social presence (i.e., the extent to which learners perceive the characters as real and relationally accessible), trust (i.e., the learners' confidence in the characters' reliability and helpfulness), and effectiveness (i.e., the perceived usefulness of the characters in supporting task completion and learning goals). Additionally, learners' intentions to adopt AI-supported learning and assessment in the future were examined. In this study, multimodal AI-generated characters—defined as characters created using Text-to-Video (TTV) technology combining synchronized visual (video) and auditory (voice) outputs —were presented to learners through pre-scripted recordings. 56 high school learners, learning English as an L2 to perform STEM practices, participated in this study. Although this study does not explore AI-generated characters that dynamically adapt to individual differences, it has implications for such contexts by highlighting how AI characters can provide personalized guidance for learning. This study contributes to AI-driven education by examining how trust and social presence shape learners' attitudes toward AI characters in learning and assessment. Accordingly, we explored the following research questions:
- To what extent do Korean high school learners perceive AI characters as facilitative of learning, in terms of their social presence, trust, and effectiveness?
- To what extent do high school learners' perceptions of AI characters' social presence, trust, and effectiveness influence their intention to adopt AI-driven assessments or instructional tools?

## 2    Related Works

### 2.1    **Scenario-based Assessment**

The increasing emphasis on 21st-century competencies has fueled a growing need for assessments that measure learners' ability to meet competency-driven collaborative goals [11]. As an assessment technique to measure complex, competency-based constructs and support diverse learner populations [2], SBA has been extensively validated across multiple disciplines, including science and L2 [11][13].

SBAs structure assessment tasks around authentic scenarios— a coherent narrative requiring learners to engage in thematically sequenced tasks that mirror expert cognitive processes and strategies. Central to these scenarios are simulated agents, who can function as mentors, offering instruction, or as collaborative partners, facilitating problem-solving. These simulated agents create a social-interactional context for authentic communication and collaboration. Unlike traditional assessments, which often avoid measuring complex competencies such as collaborative problem-solving due to concerns about reliability, SBA offers a structured yet psychometrically rigorous technique that elicits collaboration while mitigating the unpredictability of real-life interactions [11]. While simulated agents have been accounted for in SBA design in prior research [10][12], their moderating role has not been systematically validated. Furthermore, previously these agents were presented through images of static chats and voice-overs, but these methods often lacked the authenticity necessary for full learner engagement [13]. Recent advancements in multimodal AI, including Text-to-Video (TTV) technology, allow the enhancement of these pre-scripted agents into AI-generated characters capable of richer interactions [14]. However, understanding learners' perceptions of these AI characters—whether they view these agents as mentors or peers, as intended—can provide insight into their effectiveness as interactive facilitators in assessment and learning. The next section reviews research on AI characters in education.

### 2.2    **AI-generated Characters in Education**

As an attempt to foster engagement and address the diverse learning styles of learners across socioeconomic backgrounds [9], researchers have explored the use of AI-generated characters involving hyper-realistic synthesis of prose, images, audio, and video [4]. Prior research mainly explored how diverse aspects of the voices of the AI characters, generated with text-to-sound technology, influenced learners' learning experiences. Edwards et al. [5] examined how learners' perceptions of AI voice-generated virtual instructors varied by age, while Kim et al. [6] explored how different communication styles (i.e., functional vs. relational) influenced the levels of social presence.

Building on the advancement in multimodal AI, recent studies have e text-to-video, Pataranutaporn et al. [4] explored how familiarity with AI-generated instructors (e.g., historical figures vs. unfamiliar avatars) impacts learning, motivation, and attitudes. Their study underscores the importance of traceability and ethical safeguards in maintaining trust and credibility in AI-generated instructors. Furthermore, they emphasize the need for ethical guidelines, such as consent, transparency, and misinformation screening, to ensure the responsible integration of AI-generated characters in education.

Building on the shift toward multimodal, socially interactive AI characters, our study explores learners' perceptions of AI-generated characters in learning, with a focus on trust and social presence. As prior research highlights the importance of communication style, familiarity, and ethical considerations in AI-driven education, our study aims to further investigate how these factors influence learners' engagement and willingness to adopt multimodal AI characters as learning companions.

### 2.3   Trust in AI-based Educational Platforms

Trust is a key driver of technology acceptance and adoption [23], playing a critical role in AI-based educational systems as it directly influences user acceptance and engagement [15]. Qin et al. [15] identified major categories of trust-related factors in AI-based education: (a) Technology-related factors, including functionality, helpfulness, interpretability, dependability, and interaction interface, and (b) individual-related factors, such as perceptions of learning, willingness to interact with teachers, views on AI, and autonomy orientation. These factors highlight the multifaceted nature of trust in AI education, extending technical reliability to include user perceptions and interactions. The importance of learners' perceived trust in AI education is further supported by Kim et al. [7], who found that higher trust in an AI instructor increases the likelihood of learners enrolling in AI-based courses. Similarly, Amoozadeh et al. [16] demonstrate learners with higher trust in GenAI reported greater motivation and confidence in their coursework. Building on these insights, our study explores whether trust serves as a fundamental factor in shaping learners' perceptions of AI characters as learning companions, ultimately influencing engagement, adoption intentions, and learning outcomes.

### 2.4   Social Presence in AI-based Educational Platforms

Although definitions of social presence vary, most literature defines it as the perception of being connected with another entity in a mediated environment, where users temporarily overlook the presence of the medium itself [7][17]. According to Gambino et al. [18], social presence is machines exhibiting more human-like behaviors or appearances that lead users to engage with them similarly to human relationships. In the context of AI-generated pedagogical agents in K-12 learning, social presence is crucial for creating an engaging and interactive learning environment as young learners can benefit from human-like interactions and emotional connection with the AI-generated characters [19]. Specifically, studies show that social presence plays a crucial role in relationship-building with AI instructors [7], and enhancing social presence has been linked to greater emotional engagement in human-AI interactions in teacher education [8]. Building on, our study investigates how social presence impacts learners' experiences with AI-generated characters and explores its influence on the intention to adopt AI characters.

## 3     Methods

This study employed an explanatory mixed-method associational design [20] to investigate L2 learners' perceptions of AI characters in an SBA and their intention to adopt future AI-driven assessments or instructional tools. This research employed a two-phase approach: (1) a Likert scale questionnaire within an SBA to quantify learners' perceptions and attitudes, and (2) open-ended survey analysis to deepen insights and validate patterns from the quantitative data. The survey examined how learners engaged with AI characters within the SBLA-SP.

### 3.1     **Instrument: SBLA-SP**

The instrument was the scenario-based language assessment for science purposes (SBLA-SP), designed to measure specific purpose language ability in the context of performing a scientific investigation. The SBLA-SP is designed to engage learners as part of a group solving a problem with AI characters, mimicking a collaborative process through a scenario aligned with real-life science practices— specifically, planning and carrying out investigations. Learners progress through a scenario narrative, designed to elicit the five key steps of scientific investigation: (1) asking questions, (2) generating hypotheses, (3) designing experiments, (4) collecting and recording data, and (5) analyzing and interpreting data. These steps culminate in the final task: writing a lab report. The SBLA-SP was developed and delivered via Qualtrics.

The AI characters in SBLA-SP serve as mentors and teammates guiding learners through the scenario narrative. The SBLA-SP situates the learners in an online science internship designed for high schoolers. The AI characters include the mentor, Ms. Rosie, guiding the internship, and teammates Anika and Jimmy, collaborating on the science investigation project (Fig. 1). This internship takes place in a virtual science lab, where learners research triboelectric nanogenerators (TENGs)—an energy-harvesting technology that utilizes electric interactions.

AI character videos were created on Synthesia (synthesia.io), a GenAI TTV platform, and embedded into Qualtrics using HTML. Among the repository of available characters and voices in Synthesia, selections were made to ensure that each AI character was representative of their assigned role within the scenario. For example, to reflect the international internship context, characters acting as peers were carefully selected to represent high schoolers of different nationalities and accents. Given that the test was designed for L2 learners, all video outputs were reviewed using an online CEFR analyzer from Cathoven Language Hub (https://nexthub.cathoven.com/cefr) to ensure that the speech was appropriately aligned with the target CEFR B1 (intermediate-low) level for listenability. The AI characters scaffolded lab report writing and facilitated scientific investigations by situating learners in a simulated collaboration through pre-scripted videos, where they co-planned an investigation and co-constructed a lab report. Also, they provided corrective feedback, which was pre-scripted and triggered by learners' selections on multiple-choice questions, offering personalized guidance to support the scientific investigation process (Fig. 1).

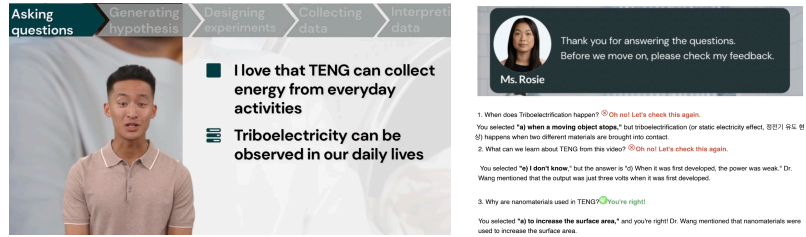

*Note:* A sample video of the characters can be found at: https://youtu.be/itwGiUIJI2Q
**Fig. 1.** AI Peer Guiding the Scenario Narrative (Left), Personalized Corrective Feedback Provided by AI Characters (Right)

Embedded within the SBLA-SP is the perception survey to capture learners' perceptions of the AI characters (Table 1). Each dimension is assessed through 4-point Likert-scale items, adapted from established sources, with a total of 35 items. The survey also includes two open-ended questions for qualitative insights.

**Table 1.** Taxonomy of the SBLA-SP Perception Survey by Construct (k = 37)

| Construct | Scale | Source | k |
| --- | --- | --- | --- |
| Tech Self-efficacy | To what extent are you open to adopting new technologies and tools (e.g., AI characters leading educational activities)? | Nass et al. [25] | 3 |
| Social Presence | To what extent did you feel like you were with the mentors and teammates? | Kim et al. [7] | 8 |
| Trust | To what extent did you trust the mentors, teammates, and their feedback? | Kim et al. [7] | 12 |
| Effectiveness | To what extent did the mentors and teammates help you learn? | Kim et al. [7] | 8 |
| Intention to Adopt | To what extent are you willing to take online classes or assessments involving AI characters? | Choi & Ji [24] | 4 |
| Total Likert Scale Items | | | 35 |
| Briefly describe how you feel about the mentor (Ms. Rosie) and her feedback | | | 1 |
| Briefly describe how you feel about the teammates (Anika & Jimmy) and their feedback | | | 1 |
| Total including Open-Ended Items | | | 37 |

### 3.2 Participants

As the SBLA-SP was designed to assess specific purpose language ability within the context of performing STEM tasks, 56 L2 learners pursuing the STEM track were recruited from high schools in South Korea, where English is taught as a foreign language through purposeful sampling [20].

### 3.3 Data Collection

The SBLA-SP was administered online via Qualtrics during a regular English class, with the teacher proctoring and addressing technical issues, while the researcher remotely monitored and assisted as needed. Before the test, the teacher informed learners that the assessment included simulated agents generated using AI technology, ensuring transparency about its use. The test was conducted on the Chrome browser, with learners using a computer, keyboard, and headphones. learners were instructed to review and electronically consent to participate in the study. They completed the test within an hour, and the assessment platform automatically recorded their responses. The perception survey was embedded within the SBLA-SP and administered at three points during the test: before the scenario began, during the scenario following feedback, and after the test concluded. Also, the sample size (N = 56) was sufficient to meet PLS-SEM's 10-times rule.

### 3.4    Data Analysis

**RQ1. CTT Investigating the Perception of AI Characters.** To investigate Korean high schoolers' perceptions of AI characters as facilitative of learning (RQ1), the classical test theory (CTT) analyses followed by qualitative analyses were conducted. First, the descriptive statistics were analyzed on Likert scale responses across three key dimensions: social presence, trust, and effectiveness. These quantitative findings were then complemented and elaborated through thematic analysis of two open-ended survey questions that queried learners' experience with AI characters.

**RQ2. SEM Assessing Learners' Intention to Adopt AI Characters.** Building on Kim et al.'s [7] model examining how learners' perceptions of an AI instructor's social presence and trust influence their intention to adopt AI character-led education, we built an exploratory model to investigate how trust and social presence impact perceived usefulness and, in turn, influence learners' behavioral intention to adopt AI characters for learning (Fig. 3). To mitigate potential confounding effects from learners' baseline technological confidence, we controlled for technology self-efficacy. The initial model was optimized and the direct influence as well as the interaction of these constructs on the behavioral intention to adopt the AI characters for learning were analyzed utilizing the Partial Least Square Structural Equation Modeling (PLS-SEM) using SmartPLS v.3.3.2.

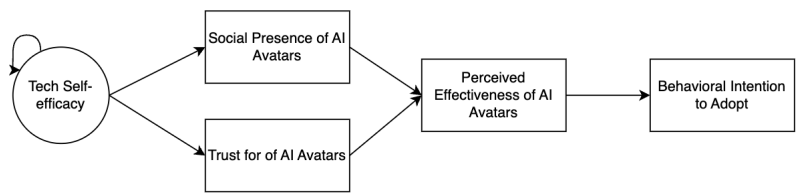

**Fig. 3.** *Initial exploratory model with variables used in the study*

PLS-SEM was chosen for its ability to combine ordinary least squares regressions with principal components analysis, which is ideal for emerging fields requiring both exploratory and confirmatory approaches [21] along with its flexibility with small-size data and its capability for causal-predictive modeling [21]. Instead of reporting model fit, which is not suitable for assessing PLS-SEM quality [21], this study validated the optimized model by evaluating construct validity, common method bias, reliability, and convergent and discriminant validity. The structural model's robustness was examined through bootstrapping and t-tests. Technology self-efficacy was measured via the survey and entered as a covariate in the PLS-SEM model, thereby controlling its confounding influence.

## 4    Results

### 4.1    **RQ1. Perception of AI characters**

To investigate the extent to which Korean high school learners perceive AI characters as facilitative of learning the survey results were analyzed. The reliability analysis demonstrated excellent internal consistency across all constructs (overall α = .96).

Analysis of the 4-point Likert scale survey revealed positive attitudes across all measured dimensions. Learners' perceptions of AI characters' effectiveness in facilitating learning were notably positive (M = 3.54, SD = 0.45), with responses trending toward strong agreement. This was closely followed by high levels of trust in AI characters (M = 3.51, SD = 0.43), suggesting learners viewed these digital entities as reliable and trustworthy learning companions. The social presence of AI characters was also perceived positively (M = 3.30, SD = 0.70), indicating that learners somewhat agreed that AI characters could create a sense of personal connection in the learning environment. Learners demonstrated adequate tech self-efficacy (M = 3.18, SD = 0.64) and showed a positive intention to adopt AI learning (M = 3.36, SD = 0.68). All dimensions supported normal distribution while exhibiting a tendency toward positive ratings within acceptable thresholds. These findings suggest that the learners perceived AI characters as facilitative of their learning, with strong perceptions of trustworthiness, social presence, and effectiveness.

**Table 2.** Reliability Analysis and Descriptive Statistics by Construct (N = 56, k = 35)

|  | k | α | Min | Max | Mean | Median | SD |
|---|---|---|---|---|---|---|---|
| *All Likert Items* | 35 | .96 | 2.29 | 4.00 | 3.42 | 3.41 | 0.44 |
| *By Construct* | | | | | | | |
| Tech Self-efficacy | 3 | .86 | 2.00 | 4.00 | 3.18 | 3.00 | 0.64 |
| Social Presence | 8 | .96 | 1.00 | 4.00 | 3.30 | 3.38 | 0.70 |
| Trust | 12 | .89 | 2.73 | 4.00 | 3.51 | 3.55 | 0.43 |
| Effectiveness | 8 | .88 | 2.11 | 4.00 | 3.54 | 3.56 | 0.45 |
| Intention to Adopt | 4 | .96 | 1.00 | 4.00 | 3.36 | 3.25 | 0.68 |

*Note: Likert Scale (1 = Strongly Disagree, 4 = Strongly Agree)*

Also, open-ended questions were analyzed through deductive coding methods using social presence, trust, and effectiveness as pillars.

**Social Presence (Realism and Naturalness) of AI characters.** Learners perceived AI characters as socially present and realistic. Many addressed the characters as "everyone" or "team members", showing a sense of high social presence. One noted, "As for the team members, even if they are virtual, they are people learning from the same standpoint, so I felt a bit more attached to them compared to the mentor." Others highlighted the immersive nature of the experience, stating, "It felt like actually doing it together," and "It was different because everyone was involved in the experiment." These responses suggest that learners recognized AI characters as active participants, making the experience feel interactive and engaging.

While some found the characters' feedback sufficient, others expressed a need for more bidirectional interaction. One appreciated the AI mentor's feedback, saying, "The fact that I can communicate directly with my AI mentor and that I can get corrections on what I write and apply them makes a big difference." However, others found the lack of two-way communication limiting, stating, "Knowing in advance that they are a virtual character and that it's not possible to have bidirectional communication makes it hard to have human emotions like attachment." Despite this, this learner acknowledged the constructive nature of AI mentor feedback, "However, I am willing to accept the mentor's feedback because it is constructive and practically helpful, and this has led to the formation of trust in the mentor."

**Trust for AI characters.** Learners' perceptions of trust in AI characters were shaped by three key factors: trustworthy material provided by the characters, reliable explanations and feedback, and consistent accuracy in information delivery. Learners appreciated the AI characters' ability to retrieve and present relevant materials. One noted, "The materials presented by the team members helped me understand how to set up hypotheses and about triboelectricity, so I think they are trustworthy." Similarly, another highlighted the reliability of the feedback, stating, "It was good because they kindly gave trustworthy feedback." These insights suggest that learners are more likely to trust AI characters when they perceive them as accurate, consistent, and supportive in their learning process.

**Effectiveness of AI characters.** Learners found AI characters effective in providing guidance, quality feedback, and learning support, which facilitated their learning experience. They appreciated the instructions AI characters provided, helping them navigate the learning process. One highlighted the effectiveness of feedback, stating, "With the simple explanations and feedback, I was able to better understand and correct my wrong answers." Others emphasized the supportive role of AI characters, with one noting, "It was fascinating and helpful to proceed with the AI avatar that provides assistance." Additionally, AI characters were seen as helpful in problem-solving, particularly when they asked questions on behalf of learners, as one learner shared, "The way the team members asked the mentor about the difficulties I faced while solving the problem helped in resolving the issue." These insights demonstrate that learners perceive AI characters as valuable learning companions providing structured guidance, high-quality feedback, and interactive support.

### 4.2   RQ2. SEM Measuring learners' Intention to Adopt AI characters

The model validation was confirmed, with details on confirmatory factor analysis, reliability, and validity provided in Appendix A. Figure 4 shows the estimation results of the relationships between constructs in the optimized model, in which the path coefficients of the model are all statistically significant.

Unlike our initial model in Figure 3, which predicted trust and social presence as simultaneous contributors to effectiveness, the optimized model revealed that trust strongly influenced social presence ($\beta = 0.65$). This underscores that trust in AI characters is a prerequisite for enhancing their perceived social presence. Also, social presence had a very strong effect on effectiveness (β = 0.88), meaning when learners felt socially engaged with AI characters, they perceived them as highly effective. Effectiveness significantly drove the intention to adopt (β = 0.78), showing that learners were more likely to embrace AI characters if they found them useful. The moderate (social presence, $r^2$=0.44) to high $r^2$ values (effectiveness, $r^2$=0.76; intention to Adopt, $r^2$=0.65) indicated that a large proportion of the variance was accounted for by the predictors. The non-significant paths of tech self-efficacy confirmed that individual technological confidence did not influence the variables.

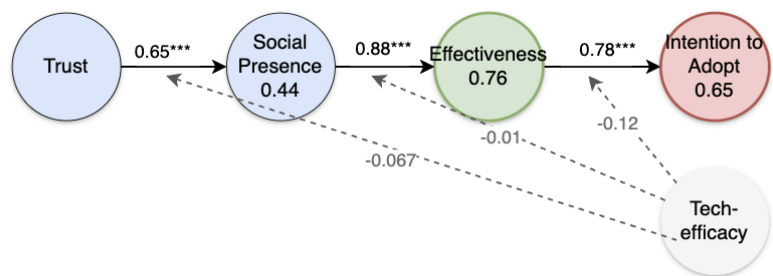

**Figure 4.** *PLS path model estimation: Bold arrowed lines show valid paths and coefficients, and dotted arrowed lines represent invalid paths. Numbers inside circles indicate $r^2$ values for constructs* (p-values * < 0.05, ** < 0.01, *** < 0.001).

In Appendix B, we report standardized coefficients for direct and indirect effects to provide a comprehensive understanding of relationships in the PLS-SEM model. In sum, the model showed that trust and social presence play a crucial role in shaping learners' perception of AI characters' effectiveness, which in turn strongly influences their intention to adopt education involving AI characters.

**Qualitative Responses on Intention to Engage with AI characters.** To further understand learners' intention to engage with AI characters, we analyzed their qualitative responses, revealing two engagement patterns: positive and conditional.

Most learners expressed enthusiasm for future adoption. One noted, "I learned about the technology TENG for the first time, and it was easy to understand as she showed and explained examples that can be found in everyday life. I look forward to activities with the mentor in the future." This suggests that the learner positively perceived the virtual mentor as an actual teaching presence showing high social presence, influenced by the usefulness of the information provided.

Others showed conditional intention, indicating that their engagement depended on improvements in AI interactions. One commented, "It was very unique and innovative….I am excited to see how much it will develop in the future. If there's a slight disappointment, I wish the feedback could be a bit more meticulous and detailed." This response refers to the AI character as "it" suggesting a lower level of social presence, and it highlights the further need for personalized feedback.

Overall, these findings indicate that while learners recognize AI characters' potential, their intention to engage largely depends on the quality of interactions, personalized feedback, and the level of social presence they experience.

## 5     Discussion

We explored learners' perceptions of multimodal AI characters, extending beyond chat-based interactions. We identified key features that enhance AI characters' effectiveness, offering two main insights for AI characters integration in education.

### 5.1     Trust as the Foundation for AI Characters in Learning

Our model analysis revealed that trust served as the prerequisite for social presence, which enhances the perceived effectiveness and willingness to adopt AI characters for learning. This finding extends Kim et al.'s [7] study demonstrating that higher trust in an AI voice was associated with a greater likelihood of enrolling in AI-based courses. Because multimodal characters combine synchronized visuals and speech, they naturally amplify social presence and raise users' expectations for human-like interaction—making trust even more pivotal than in text- or voice-only settings. We suggest that no matter how lifelike these characters appear, social presence cannot be fully realized unless trust is established. Thus, for multimodal AIs, trust is the foundation for engagement, effectiveness, and adoption.

To further explore what influences trust, we identified three key factors from the qualitative responses: (1) trustworthy material, (2) reliable explanations and feedback, and (3) accuracy in information delivery. These findings align with prior research by Amoozadeh et al. [16], which found that learners' trust in AI-generated characters declined when responses were incorrect, misleading, or unhelpful, often due to a mismatch in user expectations shaped by prior experiences with human instructors. This suggests that AI characters need to meet human-like instructional expectations to maintain trust. Additionally, Manzini et al. [22] argue that as AI assistants become more human-like, trust is further impacted by the alignment between the AI's objectives and the user's goals. Our findings support this, showing that AI characters were trusted when they provided relevant information aligned with learners' needs. In our study, AI characters had limited autonomy, only delivering human-generated instructions, feedback, and materials. Future implementations with greater AI autonomy, like real-time responses, will require strict safeguards against hallucinations through a combination of verification models, Retrieval-Augmented Generation (RAG), and a human-in-the-loop approach where humans remain accountable for source credibility [3]. Ultimately, we underscore that trust, developed through credible content and meaningful feedback, is a critical factor in AI character adoption, aligning with research on credibility in AI-driven education [23].

### 5.2     Social Presence and Collaboration with AI Characters for Learning

Our study highlights the importance of social presence, especially in a scenario simulating collaborative science investigation. When social presence was high, learners referred to AI characters as "everyone" or "team members," indicating that they viewed them as active collaborative partners in the learning process. Our learners reported experiencing stronger engagement when AI peers took the same perspective as them or acted on their behalf, mirroring human collaboration in educational settings. We found that the naturalism and realism of AI characters fostered a greater sense of social presence, which in turn enhanced perceived effectiveness. This aligns with prior research in human-AI interaction, which has established social presence as a key factor in relationship-building with AI. Machines exhibiting more human-like behaviors or appearances encourage users to engage with them similarly to human relationships [18]. Furthermore, AI systems that exhibit socially oriented behaviors, such as responsiveness combined with human-like qualities, are perceived as more capable [22]. While most prior research on AI characters has focused on their role as instructors [19], our findings emphasize the need for further investigation into AI agents that provide peer-like, empathetic interactions [7], which is particularly relevant for K-12 learners with shorter attention spans and diverse learning needs [9].

From an assessment development perspective, this study also demonstrates the potential of adopting AI characters to both foster and measure complex competencies such as collaborative problem-solving [1], which traditional assessments often avoid due to difficulties in controlling construct-irrelevant variance. The high social presence of these AI characters in our study suggests that their integration allowed structured yet dynamic social interactions that facilitate the assessment of competency-based constructs such as collaboration and critical thinking, aligning with the increasing demand for 21st-century skills [12]. By incorporating AI characters that facilitate discussion, reflection, and problem-solving, these assessments can enhance both validity and reliability while supporting the development of these crucial competencies.

In sum, our findings underscore that AI characters with high social presence not only enhance engagement and collaborative learning but also offer promising avenues for valid assessment of complex competencies, potentially transforming how we approach both instruction and evaluation in educational settings.

### 5.3  **Limitations and further study**

Several limitations of this study suggest important directions for future research. First, the AI characters in this study were pre-designed by humans, with the AI functioning only as a delivery mechanism for scripted content rather than dynamically generating responses. Future research should examine GenAI-driven interactive characters that can deliver real-time, adaptive feedback, allowing deeper insights into how trust and social presence evolve during authentic interaction. Second, while this study involved both AI mentors and AI peers, we did not conduct an extensive analysis comparing learner perceptions between these roles and did not isolate potential confounds introduced by role positions. Future studies should explore how assigned social roles influence learner perceptions independently of character behavior, using detailed qualitative analysis to disentangle potential confounds between AI capabilities and social framing. Finally, while this study captured learners' perceptions of AI character effectiveness, analyses of actual learning outcomes—defined as gains in science domain knowledge—will be reported in a future study to allow for a more comprehensive examination of the role of AI characters in promoting learning within the SBLA-SP. Taken together, these directions underscore how findings from this study can inform the thoughtful implementation of AI characters in learning and assessment, ultimately contributing to more responsive, personalized, and effective educational experiences.